\documentclass[12pt]{article}
\usepackage{ifpdf}
\pdfoutput=1
\usepackage[a4paper,width=16cm, bottom=3.2cm, left=2.5cm]{geometry}
\usepackage{fancyhdr}
\usepackage{lipsum}
\usepackage[utf8x]{inputenc}
\usepackage{graphicx,caption}
\usepackage{hyperref}
\usepackage[font=small,labelfont=bf]{caption}
\newcommand{\eq}{\begin{equation}}
\newcommand{\eqx}{\end{equation}}
\newcommand{\eqn}{\begin{eqnarray}}
\newcommand{\eqnx}{\end{eqnarray}}
\newcommand{\f}[2]{\frac{#1}{#2}}

\newcommand{\al}{\alpha}

\newcommand{\dl}{\delta}

\newcommand{\Dl}{\Delta}

\newcommand{\eps}{\varepsilon}
\newcommand{\qqqq}{\quad\quad\quad\quad}

\usepackage{microtype}

\usepackage{authblk}

\title{{\bf Conformal defects in supergravity\\
-- backreacted Dirac delta sources}}
\author{Romuald A. Janik\footnote{Email: romuald@th.if.uj.edu.pl }}
\author{Jakub Jankowski\footnote{Email: jakubj@th.if.uj.edu.pl}}
\author{Piotr Witkowski\footnote{Email: picek.witkowski@uj.edu.pl}}
\affil{Institute of Physics\\ 
Jagiellonian University\\  
ul. {\L}ojasiewicza 11\\ 
30-348  Krak{\'o}w\\ 
Poland}
\date{}

\begin{document}

\maketitle

\thispagestyle{empty}

\abstract{We construct numerically gravitational duals
of theories deformed by localized Dirac delta sources
for scalar operators both at zero and at finite temperature.
We find that requiring that the backreacted geometry preserves
the original scale invariance of the source
uniquely determines the potential
for the scalar field to be the one found in a certain Kaluza-Klein
compactification of $11D$ supergravity. 
This result is obtained using an efficient perturbative expansion
of the backreacted background at zero temperature and is confirmed
by a direct numerical computation.
Numerical solutions at finite temperatures
are obtained and a detailed discussion of the numerical approach
to the treatment of the Dirac delta sources is presented.
The physics of defect configurations is illustrated 
with a calculation of entanglement entropy.}

\newpage


\section{Introduction}

Many phenomena in condensed matter physics involve
strongly interacting systems and it is suspected that
in many cases the physics is governed by a quantum critical point, 
with effective scale invariance. Thus the description involves strongly coupled
conformal field theories (CFT).
A widely used tool to analyze certain strongly coupled CFT systems is the anti-de Sitter
conformal field theory (AdS/CFT) correspondence \cite{Maldacena:1997re,Witten:1998qj,Gubser:1998bc}, 
a setup in which a nongravitational system is mapped
to a theory with gravity in a higher dimensional spacetime.
Over the past few years, it has been applied to model many interesting 
phenomena, including superconductivity and superfluidity,
Fermi surfaces and non-Fermi liquids -- see \cite{Hartnoll:2009sz,McGreevy:2009xe} for a review.

The initial applications of AdS/CFT to model condensed matter like systems
typically involved exact translation invariance which then by construction
made some relevant physics inaccessible or obscured. 
Hence one of the main directions in current research
is to incorporate such key elements of condensed matter systems as atomic lattices and localized defects
into a dual gravitational AdS/CFT description.

There have been various ways of how to introduce lattices into the correspondence including Q-lattices 
\cite{Donos:2013eha,Andrade:2013gsa,Gouteraux:2014hca}, helical lattices \cite{Donos:2012js,Donos:2014oha}, 
single momentum modes in scalar field (neutral lattices) \cite{Horowitz:2012ky,Rangamani:2015hka} or in chemical
potential (ionic lattices) \cite{Horowitz:2013jaa,Donos:2014yya,Ling:2013nxa}.
Also very recent studies of localized charge defects \cite{Blake:2014lva,Horowitz:2014gva}
brought interesting physical insights an example being holographic Friedel charge oscillations \cite{Blake:2014lva}.
Most of these models incorporated the breaking of translation invariance through the introduction
of spatially varying \emph{smooth} sources deforming the CFT action 
typically including one to a few Fourier modes. 

A quintessential model of a solid state lattice is the venerable Kronig-Penney model \cite{KPm1931},
in which a periodic lattice of Dirac delta functions is imposed in the potential in 
the Schr{\"o}dinger equation.
This allows for quasi-analytical calculations of the key physical quantities of
interest. Our main motivation is to develop techniques for dealing with
a similar setup on the dual holographic gravity side. Although seemingly this looks like
an innocuous generalization from a single Fourier mode source to one where
all the Fourier modes are turned on in the source\footnote{To avoid confusion,
throughout this paper by `a source' we always mean the deformation of the field theory CFT
Lagrangian. Within the AdS/CFT dictionary this is encoded in the \emph{boundary conditions}
for the bulk gravitational/scalar/gauge fields. The bulk Einstein-matter equations
do not have in contrast any external sources.} in a uniform manner,
the nonlinear nature of the bulk gravitational description makes this problem
very challenging.
This is especially so as we need to employ numerical relativity methods and imposing
distributional Dirac delta like boundary conditions is very nontrivial
and we lack guidance from the conventional numerical relativity literature.
By itself this problem is thus also quite interesting purely from the numerical relativity point
of view. 

This paper is a first step in this direction where we construct a fully backreacted
gravity and scalar configuration with a single (1D) Dirac delta source both at zero
and at nonzero temperature. We will consider a periodic version in a forthcoming work. 

We should note, however, that investigations of localized configurations in holography have, despite recent applications, 
a much longer history starting with a seminal
Janus configuration of Type IIB supergravity  \cite{Bak:2003jk}.
Subsequent research include configurations with different supersymetry
breaking patterns \cite{D'Hoker:2006uu,Clark:2005te,D'Hoker:2007xy,D'Hoker:2007xz,Bobev:2013yra,Suh:2011xc}
and configurations at finite temperature \cite{Bak:2011ga}. A particulary interesting case has 
been found in the context of eleven dimensional supergravity  \cite{DHoker:2009gg}, where
localized defect solution was demonstrated for a scalar field of mass $m^2=-2$. This solution
is a one parameter, regular deformation of the  $AdS_4\times S^7$ vacuum and 
preserves half of the original supersymmetry as well as residual conformal symmetry. 
However, it contains non trivial profiles of various $p-$forms and the special 
ansatz makes it difficult for finite temperature and finite charge generalizations.

This motivates the search for cases more suited for applications.
In particular we will employ a relatively minimal framework developing
on ideas initiated in \cite{Horowitz:2012ky} where
a single Fourier mode source for the scalar field played a role of an atomic
lattice. We consider the opposite limit of exciting all modes with equal 
amplitudes. This means imposing a Dirac delta function source for the scalar field.
Surprisingly, we find that a scale invariant Dirac delta source (which occurs
in our setup for a linear 1D Dirac delta source) leads to a consistent
scale invariant backreacted background \emph{only} if the self-interaction
potential for the scalar field is exactly equal to the one appearing 
in dimensional reductions of supergravity.
In this case we can control the numerics directly for the case of a Dirac delta
source without any regularization and extend the computation to nonzero temperature.


The paper is organized as follows. In section \ref{defect} we formulate the
problem and present a linearized analysis. Section \ref{solT0} contains 
a construction of the backreacted solution at $T=0$ together with 
the unique determination of the scalar potential from the requirement
that the backreacted geometry preserves the scale invariance of the
linear Dirac delta deformation.
The finite temperature solution is constructed in section~\ref{solTnon0}
using the DeTurck method adapted to the Dirac delta asymptotics. We also
perform numerical cross checks with regularized delta-like sources.
The physics of the configuration is 
illustrated with a calculation of the entanglement entropy in 
section \ref{EE}, both for the $T=0$ and nonzero temperature
cases. We close the paper with conclusions and a summary.


\section{Defect source for the scalar field}
\label{defect}

Let us consider a general action for a
real, self-interacting
scalar field coupled to gravity 
		\begin{eqnarray}
      S = \frac{1}{16\pi G_N}\int d^4 x \sqrt{-g}\Bigg[R-\frac{1}{2}\nabla_a\phi\nabla^a\phi  - V(\phi)\Bigg]~, 
			\label{S}
    \end{eqnarray}
where we choose
\begin{equation}
 V(\phi) = -6 - \phi^2 - \sum_{k=1}^{\infty} c_k \phi^{2k+2}~,
\label{eq:Vexp}
\end{equation}
with the coefficients $c_k$ being for the moment arbitrary. With such a definition
the leading order terms give the right cosmological constant 
and determine the mass of the scalar to be $m^2=-2$.
The vacuum solution of this theory is empty $AdS_4$ space 
with a vanishing scalar. Throughout this paper we consider
the Poincare patch
\begin{equation}
	ds^2 = \frac{1}{z^2}\Bigg(dz^2 - dt^2 + dx^2 + dy^2\Bigg)~.
	\label{eq:AdS}
	\end{equation}
with $z$ being the bulk coordinate. The asymptotics of the scalar field 
near the conformal boundary ($z=0$) are
  \begin{equation}
    \phi(x,z)\sim \phi_1(x)z+\phi_2(x)z^2~.
  \label{eq:phiasympt}
  \end{equation}
According to the standard holographic dictionary $m^2 = \Delta(\Delta-3)$, and hence
we have two allowed solutions
$\Delta=2$ and $\Delta=1$. Both of the choices are possible and
we choose to set
$\phi_1(x)$ as a source of 
an operator $\mathcal{O}(x)$ of conformal dimension $\Delta=2$. Then the subleading term is related
to the corresponding expectation value $\phi_2(x)=\langle\mathcal{O}(x)\rangle$.


\subsection{Linearized analysis}

As a first step of the analysis it is instructive to impose a 1D Dirac delta
source
\eq
\label{e.phi1}
\phi_1(t,x,y)=A_0 \, \dl(x)~,
\eqx
and find a
linearized scalar profile around empty $AdS_4$ with neglected
backreaction on the geometry.
The solution which has the correct boundary conditions 
for a delta function located on a line $x=0$ is 
   \begin{equation}
     \phi_{\rm lin}(x,z) = \frac{A_0 z^2}{\pi(x^2+z^2)}~,
   \label{eq:P0}
   \end{equation}
which is essentially just a bulk-boundary propagator.
  On the dual theory side this corresponds to the shift of the original 
	lagrangian
	\begin{equation}
	\mathcal{S}= \mathcal{S}_{\rm CFT_3} + \int d^3 x~\phi_1(x)\,\mathcal{O}(x)~,
	\label{eq:L}
	\end{equation}
	%
	which induces the vacuum expectation value $\langle\mathcal{O}(x)\rangle\sim1/x^2$. 
	This approximation is valid for $A_0<<1$.
	
	This deformation plainly breaks the translational invariance of the theory. The original $SO(3,2)$ 
	conformal symmetry is broken to the $SO(2,2)$  conformal symmetry of $d=(1+1)$ dimensions along
	the defect, where the operator is sourced. This is easily seen when one remembers that 1D Dirac delta 
	has scaling dimension equal to 1, which, together with scaling of the sourced operator, exactly cancels 
	scaling dimension of the integration measure. In order to make this symmetry manifest we can adopt
	the $AdS_3$ slicing coordinates \cite{Bak:2003jk,DHoker:2009gg} in which the background metric takes the following form 
			\begin{equation}
			ds^2 = \frac{1}{\cos(\alpha)^2}\Bigg(d\alpha^2 + \frac{dr^2-dt^2+dy^2}{r^2}\Bigg)~.
			\label{eq:aAds}
			\end{equation}
     In these coordinates linearized fluctuations around (\ref{eq:aAds}) are governed by the equation
     \begin{equation}
     \frac{d^2\phi_{\rm lin}}{d\alpha^2}+2\tan(\alpha)\frac{d\phi_{\rm lin}}{d\alpha}+\frac{2}{\cos(\alpha)^2}\phi_{\rm lin} = 0~,
     \label{eq:linphi}
     \end{equation}
	with the solution possessing the right boundary condition being of the simple shape
	\begin{equation}
	\phi_{\rm lin}(\alpha) = \frac{A_0}{\pi}\cos(\alpha)^2~.
	\label{sol:philin}
   \end{equation}	 
   It is easy to see that the coordinate change $z=r\cos(\alpha)$ and $x=r\sin(\alpha)$ 
   this solution
   transforms back to ordinary Poincare coordinates
   (\ref{eq:P0}).


\section{Backreacted solution at $T=0$}
\label{solT0}

In this section we will construct a fully backreacted solution for the one dimensional
delta-like defect (\ref{e.phi1}) at zero temperature $T=0$. We will
require the solution to possess the residual scaling symmetry
of the linearized case.
      In order to solve the full set of Einstein-scalar equation we 
			adopt the $AdS_3$ slicing coordinates familiar from the consideration of Janus solutions with a minor modification
			\begin{equation}
			ds^2 = \frac{1}{A(\alpha)^2}\Bigg(\frac{d\alpha^2}{p^2} + \frac{dr^2-dt^2+dy^2}{r^2}\Bigg)~.
			\label{eq:Aads}
			\end{equation}
Here $p$ is a constant introduced in order for the $AdS$ boundary to be located always at $\al=\pi/2$. This is
ensured by imposing $A(\pi/2)=0$ as a boundary condition.
			In this coordinate system $AdS$ space with linearized scalar profile (\ref{eq:P0}) is
			\begin{equation}
			A(\alpha)=\cos(\alpha)~,
			\hspace{1cm}
			p=1~,
			\hspace{1cm}
			\phi_{\rm lin}(\alpha)=\frac{A_0}{\pi}\cos(\alpha)^2~.
			\label{eq:8}
			\end{equation}
			The conformal
			boundary consists of two parts $\alpha_0=\pm\pi/2$ joined together along the defect.
			Transformation to the Poincare coordinates is in that case $z=r\cos(\alpha)$ and $x=r\sin(\alpha)$
			and will get modified in the full backreacted solution.
			Due to the symmetry
			of the problem all the relevant fields depend only on $\alpha$.
			The coupled set of Einstein-scalar equations reads
			\begin{equation}
			R_{ab} - \frac{1}{2}\left(\nabla_a\phi\nabla_b\phi + g_{ab}V(\phi)\right) =0~,
			\end{equation}
			\begin{equation}
			\nabla_a\nabla^a\phi-\frac{d V}{d\phi}=0~,
			\end{equation}
			which in our case explicitly gives
			\begin{equation}
			-V(\phi (\alpha ))-p^2 \left(A(\alpha ) \left(A(\alpha ) \phi '(\alpha )^2-6 A''(\alpha )\right)+6 A'(\alpha )^2\right)=0~,
			\label{eq:E1}
			\end{equation}
			\begin{equation}
		  p^2 A(\alpha ) A''(\alpha )-3 p^2 A'(\alpha )^2-2 A(\alpha )^2-\frac{1}{2} V(\phi (\alpha )) = 0~,
			\label{eq:E2}
			\end{equation}
			\begin{equation}
			p^2 A(\alpha ) \left(A(\alpha ) \phi ''(\alpha )-2 \phi '(\alpha ) A'(\alpha )\right)-V'(\phi (\alpha )) = 0~.
			\label{eq:E3}
			\end{equation}
			From the first two of the above equations we can obtain a first order ordinary differential
			equation for the function $A(\alpha)$ convenient for numerical or perturbative analysis
			\begin{equation}
			6 p^2 A'(\alpha )^2+A(\alpha )^2 \left(6-\frac{1}{2} p^2 \phi '(\alpha )^2\right)+V(\phi (\alpha )) = 0~.
			\end{equation}
It is instructive to first perform a perturbative expansion of the solution 
			\begin{equation}
			A(\alpha) = \sum_{n=0}^\infty A_n(\alpha)\epsilon^{2n} ~,
			\hspace{13pt}
			\phi(\alpha)= \sum_{n=0}^\infty f_n(\alpha) \epsilon^{2n+1}~,
			\hspace{13pt}
			p = \sum_{n=0}^\infty p_n \epsilon^{2n},
			\end{equation}
			where the lowest order is the $AdS$ solution with the scalar profile (\ref{eq:8}).
It is convenient to identify the expansion parameter $\epsilon$ with the value of the
scalar field at $\al=0$.
			
We demand the following boundary conditions for equations (\ref{eq:E1})-(\ref{eq:E3}):
we assume reflection symmetry around $x=0$ which implies
\eq
\partial_\alpha A_n(0) = 0~,   \qqqq \partial_\alpha f_n(0) = 0~.
\eqx
For the scalar field in addition we require $f_n(0)=0$, which ensures that  
$\epsilon$ becomes a physical expansion parameter i.e. it remains equal to
the value of the scalar field at $\al=0$ at any order in the perturbative expansion. 
We determine the constants $p_n$ by requiring that the $AdS$ boundary is always at $\al=\pi/2$
through $A_n(\pi/2)=0$.

The above conditions, for a given choice of the scalar potential (\ref{eq:Vexp})
determine a \emph{unique} solution. A surprising generic feature of the obtained solution
is that $f_n'\left(\frac{\pi}{2}\right) \neq 0$, which, when translated to
standard Fefferman-Graham coordinates leads to a nonvanishing nonnormalizable mode away from $x=0$
i.e. a modification of the original Dirac delta source to 
\eq
\phi_1(x) =\eps \delta(x) + (\eps^3+\ldots)\frac{1}{|x|}~.
\eqx
Since we want to have a purely localized Dirac delta source, we impose the additional
condition
\eq
			f_n'\left(\frac{\pi}{2}\right)=0~,
			\label{eq:cond}
\eqx
as an equation for the coefficients of the scalar potential (\ref{eq:Vexp}) which turn out to be
uniquely fixed order by order. The first couple of coefficients are $c_1=1/36,~ \hspace{3pt} c_2=1/3240,~  
c_3=1/544320,~c_4=1/146966400,~\hspace{3pt} c_5=1/58198694400$.
Those are exactly the same as the first terms in the Taylor series expansion of
			\begin{equation}
			V(\phi) = -6\cosh\left(\frac{\phi}{\sqrt{3}}\right)~.
			\label{eq:Vsugra}
           \end{equation}					
Consequently, only for this potential does a backreacted \emph{conformal} defect with a
Dirac delta function source on a line exist.

It is interesting to note that the above potential is not accidental and comes from 
a certain Kaluza-Klein compactification of $D=11$ supergravity
after truncation of equations of motion to $\mathcal{N}=2$
supersymmetry \cite{Cvetic:1999xp,Duff:1999gh} 
(see \cite{Duff:1999rk} for a review).
The minimal lagrangian of such a reduction, apart from the
scalar and the graviton, contains one $U(1)$ gauge field coupled
in a non-minimal way with the scalar field
\begin{eqnarray}
      S = \frac{1}{16\pi G_N}\int d^4 x \sqrt{-g}\Bigg[R-\frac{3}{4}e^{\phi/\sqrt{3}}F_{ab}F^{ab}-\frac{1}{2}\nabla_a\phi\nabla^a\phi 
+ 6\cosh\left(\frac{\phi}{\sqrt{3}}\right) \Bigg]~. 
\label{Sugra}
\end{eqnarray}

\begin{figure}[th]
\begin{center}
\includegraphics[height = .205\textheight]{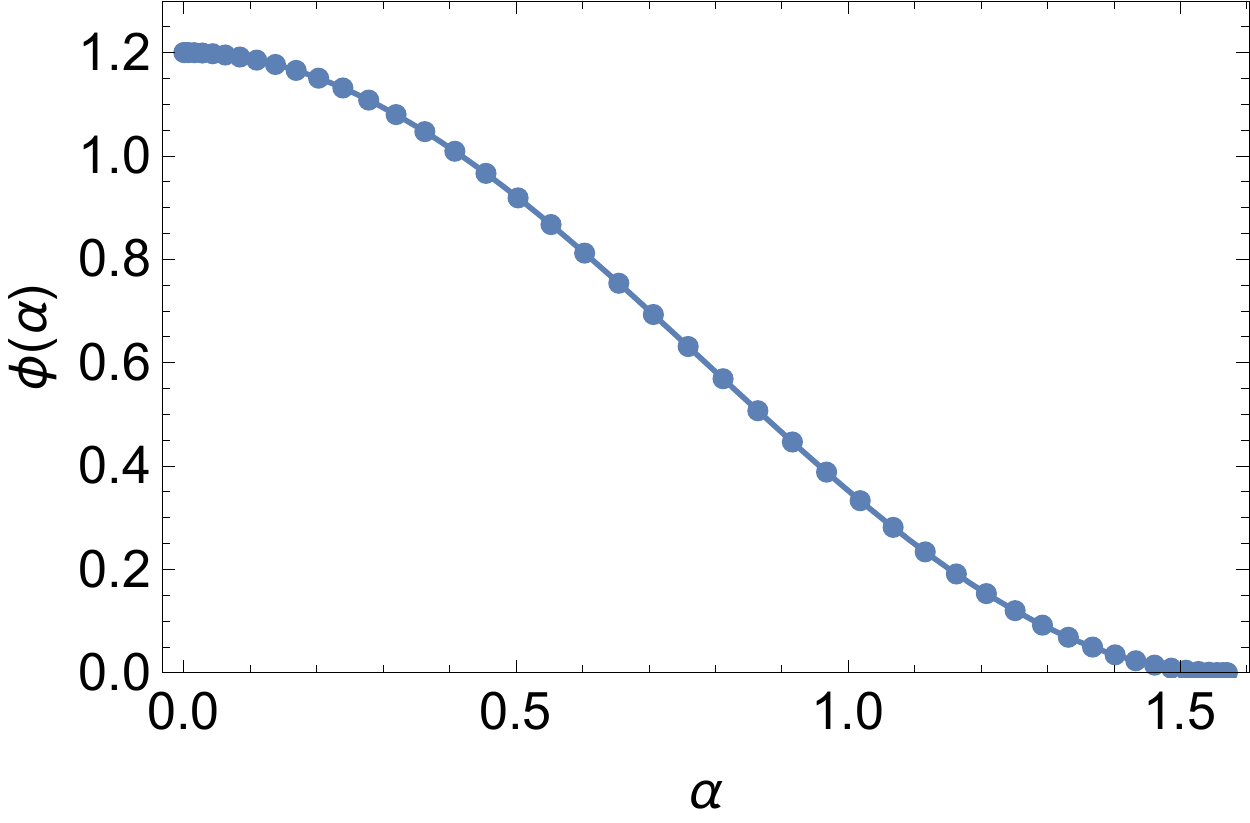} 
\includegraphics[height = .21\textheight]{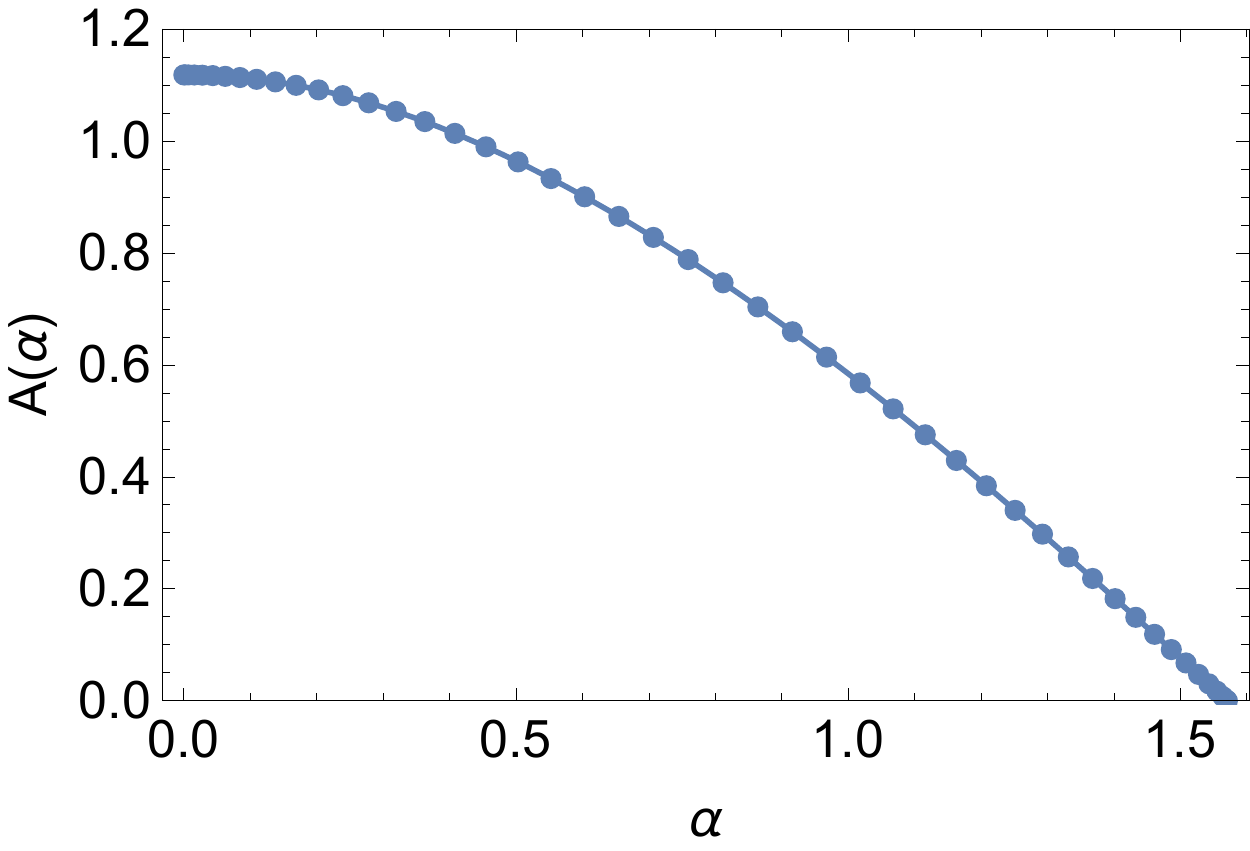}
\caption{Metric and scalar field for $\phi(0)=1.2$.
Numerical solution (points)
with $N=47$ spectral grid.
Lines   correspond to fourth order perturbative
solution.}
\label{gT0}
\end{center}
\end{figure}

The result that for the conformal defect to exist the scalar self interaction potential
has to be of the SUGRA form is indeed quite surprising at first sight.
This may be qualitatively understood in analogy to deforming a CFT by marginal
or exactly marginal operators. In the former case, on the linearized level
one still has a CFT, but if the operator is not exactly marginal, at higher orders in
the deformation parameter a mass scale is generated and the deformed theory looses
scale invariance. The phenomenon that we are seeing here is analogous but for
a linear 1D Dirac Delta source. On the linearized level we have scale invariance
just by dimensional analysis. However we may expect anomalous scaling on the
fully nonlinear level w.r.t. the deformation parameter. The fact that this does not happen for a theory with
a supergravity dual is quite natural as we may expect that for a supersymmetric
field theory there may be appropriate cancellations which would ensure
the `exact marginality' of the defect deformation.
However it would be very interesting to understand this in more detail.

As a cross check of the perturbative considerations we implemented the 
system (\ref{eq:E1})-(\ref{eq:E3}) numerically for the 
specific supergravity choice of potential (\ref{eq:Vsugra})
and checked that we can obtain a consistent backreacted geometry
with the purely localized Dirac delta source (i.e. satisfying $\partial_\al \phi(\pi/2)=0$)
for finite values of $\phi(0)$. It is at this stage that the introduction
of the constant $p$ was particularly useful as it made the size of the numerical
grid to be fixed and the same irrespective of the value of $\phi(0)$.
  
	                For numerical simulations we used standard spectral collocation method
	                \cite{Grandclement:2007sb}
					with Chebyshev polynomials to account for the $\alpha$ dependence
					and solving the resulting non-linear algebraic equations by the Newton-Raphson method.
					Curves on the plots in Fig. \ref{gT0} were calculated  with $N=47$ spectral points.


\section{The finite temperature solution}
\label{solTnon0}

In view of possible applications it is interesting and natural to generalize the configurations from
the previous section to the finite temperature case.
This will no longer 
reduce to a system of ordinary differential equations and will 
depend on both variables $\alpha$ and $z$ making the problem
much more involved. We will employ the DeTurck method with
appropriate modifications for incorporating 
the Dirac delta source.

\subsection{The DeTurck method}

The DeTurck method amounts to adding to the original
Einstein equations carefully chosen terms that make them 
elliptic partial differential equations. One solves the resulting equations
numerically and then makes sure that
this solution solves the initial problem with the original Einstein equations. 
It was first used to prove the short time existence of solutions
of the Ricci flow equation \cite{DT2003}.
In the context of finding numerically static black hole solutions
it was used in references \cite{Wiseman:2011by,Headrick:2009pv}.

In the general case the equations of motion take the following form
\begin{equation}
G_{ab}= R_{ab} - \frac{1}{2}\Bigg(\nabla_a\phi\nabla_b\phi + V(\phi)g_{ab}\Bigg)  = 0~,
\end{equation}
\begin{equation}
\nabla_a\nabla^a\phi-\frac{d V}{d\phi}=0~.
\end{equation}

\begin{equation}
G_{ab} - \nabla_{(a}\xi_{b)} = 0~
\label{eq:DT}
\end{equation} 
where 
\begin{equation}
\nonumber
\xi^a = g^{cd}[\Gamma_{cd}^a - \bar{\Gamma}_{cd}^a]~
\label{eq:Xi}
\end{equation}
and $\bar{\Gamma}(\bar{g})$ are Christoffel symbols
of the reference metric, which we take to be the standard black
hole metric.

The generic ansatz for the metric we take is
\begin{eqnarray}
ds^2  &=& \frac{1}{z^2}\Bigg[-(1-z)G(z)H_1(x,z)dt^2 + \frac{H_2(x,z)dz^2}{(1-z)G(z)}\\
\nonumber
&+& S_1(x,z)(dx + F(x,z)dz)^2 + S_2(x,z)dy^2\Bigg]~, 
\label{eq:Metric}
\end{eqnarray}
where we have set AdS radius to $L=1$ and
factored out
\begin{equation}
G(z) = 1+z+z^2~.
\label{eq:G0}
\end{equation}
With proper regularity conditions at
$z=1$ this geometry will have a smooth 
horizon with the temperature
$T = \frac{G(1)}{4\pi} = \frac{6}{8\pi}$.
This ansatz was used in the study of \cite{Horowitz:2012ky} where the boundary
source was a single Fourier mode $\phi_1(x) \propto \cos k x$.
For the case of a Dirac delta source $\phi_1(x) \propto \dl(x)$,
we need to judiciously modify the coordinate system in order to take
into account the high variability of the metric coefficients close
to Dirac delta source at the boundary $x=z=0$.
Note that in the DeTurck method, since the ansatz for the metric
is always the most general, the change of the coordinate system
essentially amounts to an appropriate modification of the reference metric. 

Motivated by the treatment of the $T=0$ case we define
\begin{equation}
\tan(\alpha) = \frac{x}{z}~.
\label{eq:BHangleVariable}
\end{equation}
keeping the second relevant coordinate $z$ unmodified.
In contrast to the $T=0$ case the geometry will depend on both
variables $(z,\alpha)$ and at $z=1$ we assume the appearance of 
a regular, static event horizon.


\subsection{The boundary ODE system}

In the new coordinate
system the two sides of the boundary on both sides of the 
defect are represented by two points $z=0$ and 
$\alpha=\pm\pi/2$. The unknown functions on an
open interval $\{0\}\times(-\pi/2,\pi/2)$
represent the (backreacted) infinitesimal neighbourhood of the delta source and thus have to 
be determined from the equations of motion.
The resulting solution will then provide the 
right boundary conditions at $z=0$ for subsequently solving the DeTurck
equations in the bulk.
This is the major necessary modification of the standard setup,
where typically the boundary conditions at $z=0$ are completely trivial
and explicitly known from the outset.

To solve the above problem we expand the 
equations of motion near the $z=0$ point and take 
the leading order terms, which provide 
a \emph{closed} self-consistent\footnote{In particular no $z$ derivatives appear.} 
set of coupled, second order ordinary differential equations
for the boundary values of the fields
which we then numerically solve. For $\alpha=0$ we set
symmetric boundary condition i.e. we set all functions,
except for $F(\alpha,0)$, to be symmetric. 
The off-diagonal function has to be clearly anti-symmetric.
For $\alpha=\pi/2$ we
choose empty $AdS$ boundary condition with a vanishing scalar.
The obtained solution at $z=0$ will be used as a 
Dirichlet boundary condition at the conformal boundary
for the set of equations in the bulk.


\subsection{The backreacted geometry for $T\neq 0$}

\begin{figure}[ht]
\begin{center}
\includegraphics[height = .25\textheight]{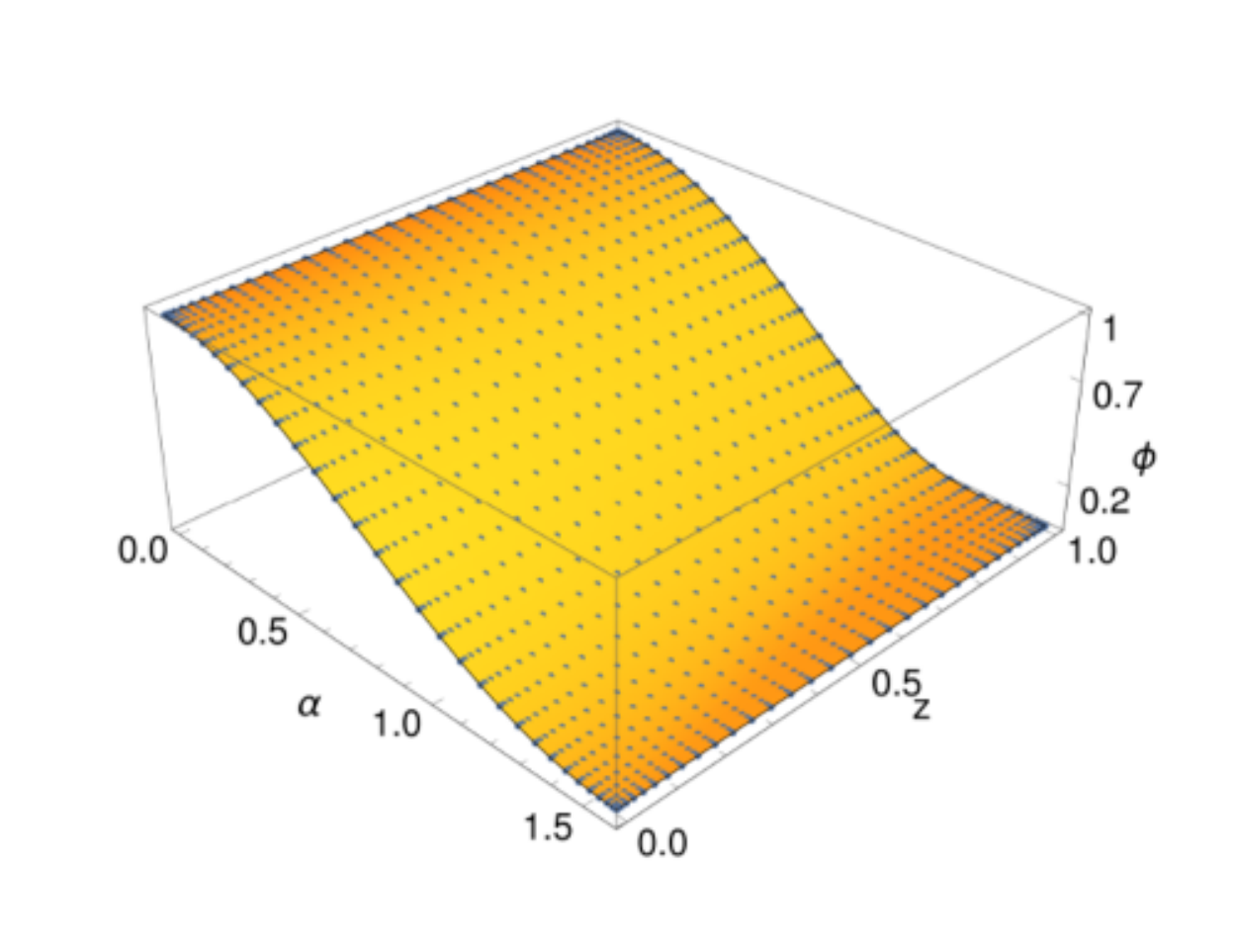} 
\includegraphics[height = .25\textheight]{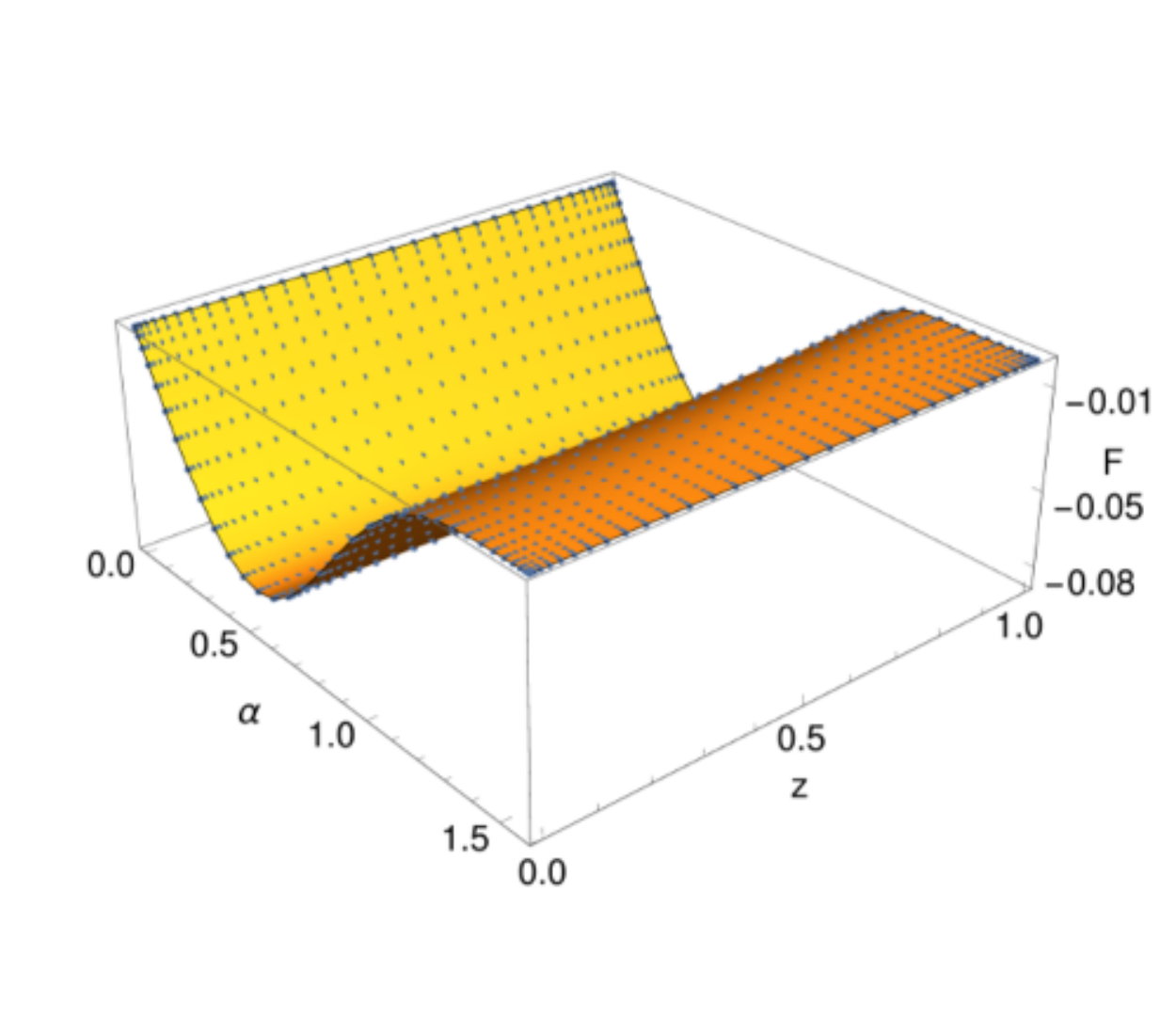}
\caption{Metric and scalar field for $\phi(0)=1.0$.
Numerical solution (points)
with $N_\alpha=N_z=35$ spectral grid.}
\label{DeltaT}
\end{center}
\end{figure}

As was discussed in the previous subsection we 
assume symmetry in $\alpha$ variable and use 
solution of the boundary ODE system as a Dirichlet
condition for $z=0$. The line $\alpha=\pi/2$ 
corresponds to the line $x=\infty$. We set 
there $AdS$ black hole condition with a 
vanishing scalar field. At the horizon $z=1$
we impose the regularity of the solution following
\cite{Horowitz:2012ky}.
As in the previous case of zero temperature
we use the same spectral
methods \cite{Grandclement:2007sb}
combined with a {\tt scipy} nonlinear solver in Python.
Sample solutions for $(\alpha,z) \in[0,\pi/2] \times [0,1]$
are shown in the plot on Fig. \ref{DeltaT}. We see that the
special choice of coordinate system rendered
the Dirac delta source well behaved with the solution being
regular and showing no anomalies which are present in the case without supersymmetric
potential. The geometry smoothly interpolates between the horizon and the
delta defect located at the boundary. Surprisingly there is only little dependence
on the holographic direction as one goes from the UV to IR where the 
horizon is located.


While solving (\ref{eq:DT}) numerically we have to make
sure that the solution is also a solution to the original Einstein
equations. In the pure gravity case one can show that it
is indeed true and no Ricci solitons are possible
\cite{Wiseman:2011by}.
In the case with matter this question remains open and 
we checked numerically, that our solutions indeed have $\xi=0$.


\subsection{Regularized Dirac delta at $T\neq 0$}

To cross-check our numerics we also constructed backreacted geometries
corresponding to a set of regularized
localized defects, which converge to an exact Dirac delta for
a vanishing regulator. 
In this case we take as a quasi-localized source for the scalar 
\begin{equation}
\phi_1(x) = \frac{z_0}{x^2+z_0^2}~,
\end{equation}
which is
motivated by the linear solution of our problem.
Here $z_0$ determines the width of the configuration
and in the limit $z_0\rightarrow0$ we recover the strict
Dirac delta source of the previous sections.

In this case we use a modified version of our coordinates
namely we define
\begin{equation}
\tan s = \frac{x}{z+z_0}~,
\label{eq:trans}
\end{equation}
so that our source term takes the form
\begin{equation}
\phi_1(s) = \frac{1}{z_0(\tan^2s+1)}~.
\label{Source}
\end{equation}
As previously $s$ is in the range $s\in[0,\pi/2]$.

The treatment of boundary conditions is similar to the
previous case but for the significant simplification
that for the $z=0$
conformal boundary we can now directly impose the source
(\ref{Source}) for the scalar and empty $AdS$
metric for the remaining functions.


\begin{figure}[th]
\begin{center}
\includegraphics[height = .21\textheight]{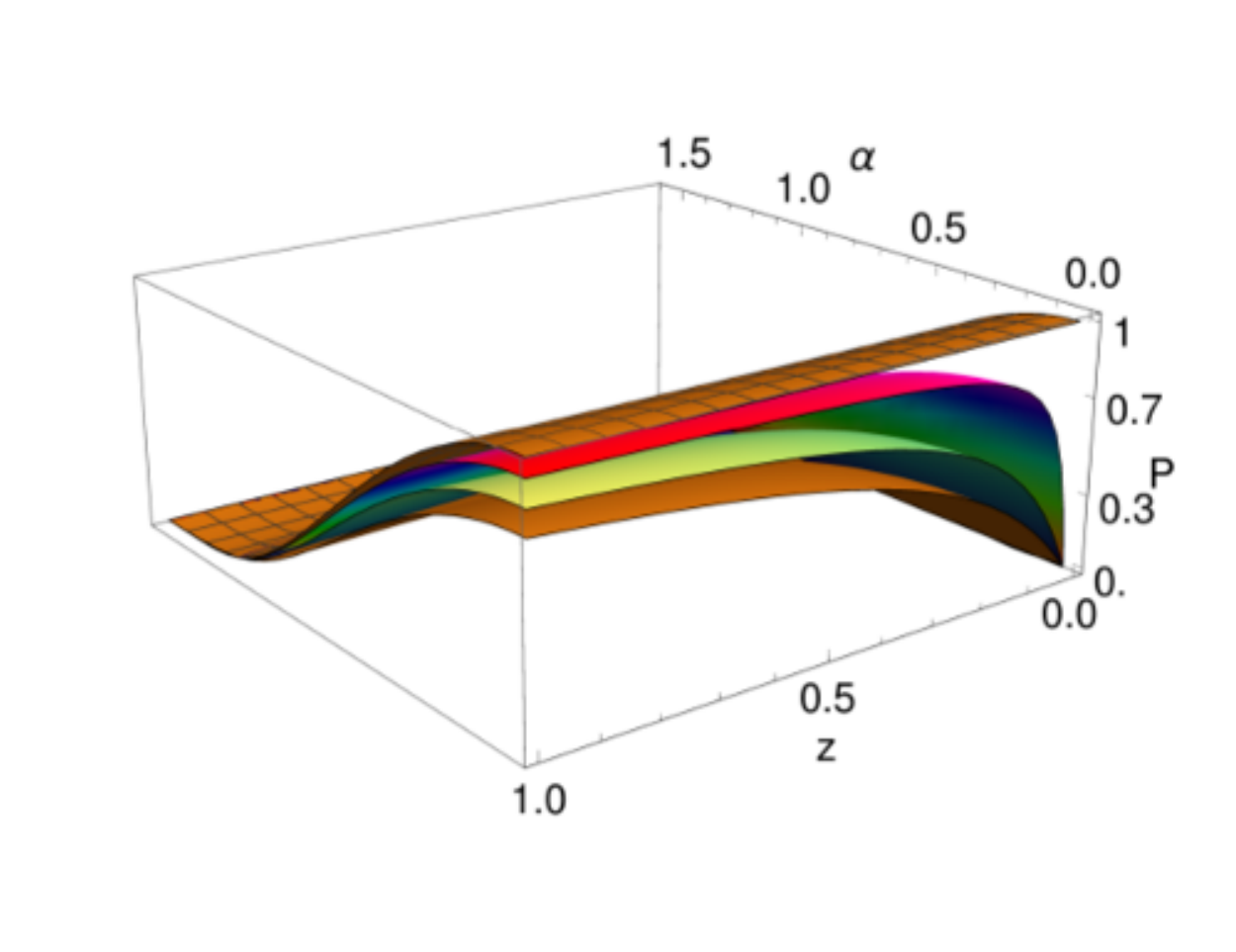} 
\includegraphics[height = .21\textheight]{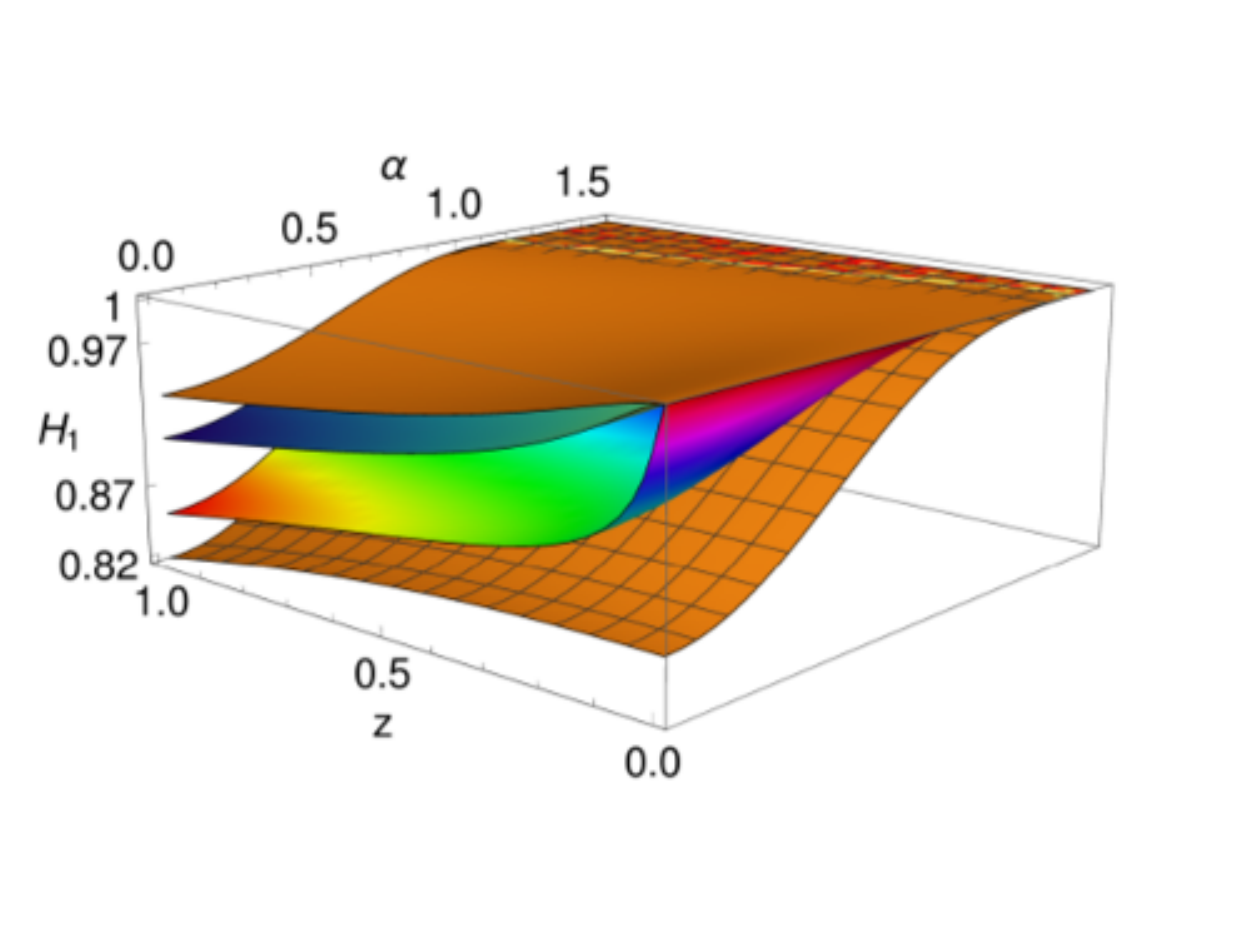}
\caption{Metric and scalar field for $\phi(0)=1.0$.
Numerical solution (points)
with $N_s=N_z=35$ spectral grid. The checkered surface is
the solution for the exact Dirac delta, while the remaining three
surfaces correspond to regularized defects with the regulator being
$z_0=0.3, 0.15, 0.025$. }
\label{Conv}
\end{center}
\end{figure}


In order to compare the regularized solutions with
the backreacted geometries from the previous sections
we need to match the relevant parameters which is not
completely trivial.
The solutions of the exact defect are parameterized by
the value of the scalar field at the point $s=0$ and $z=0$.
One recovers the amplitude by an integration of the source 
term over the conformal boundary
\begin{equation}
A_0 = \lim_{z\rightarrow0} \int_{-\infty}^{\infty} dx \frac{1}{\pi z}\phi(x,z)
=
 \lim_{z\rightarrow0}\int_{-\pi/2}^{\pi/2} d\alpha  \frac{z+z_0}{\pi z\cos(\alpha)^2}\phi(\alpha,z)~,
\end{equation}
where the last equality comes from the coordinate transformation
(\ref{eq:trans})
and the $1/\pi$ term comes from normalization. This function turns out 
to have a fixed point for $\phi(0,0)=1$ i.e. for that value of the 
field $A_0=1$. In Fig. \ref{Conv} we see that for smaller and smaller values of 
the regulator, the results converge to the solution 
corresponding to the exact Dirac delta defect.


\section{Entanglement entropy}
\label{EE}

Entanglement entropy is an interesting probe of various physical systems,
for which there exists a clear holographic prescription due to Ryu and Takayanagi \cite{Ryu:2006bv,Ryu:2006ef} 
(see \cite{Nishioka:2009un} for a review).
Apart from being of direct physical interest, we will employ entanglement entropy
also as an important nontrivial cross-check of our numerical relativity
constructions of the backreacted geometries with Dirac delta sources.
Namely, we will link the small size limit of the $T\neq 0$ entanglement
entropy with the $T=0$ one. Due to the fact that the relevant $T\neq 0$ and $T=0$
background geometries were obtained using quite different methods in qualitatively
different coordinate systems, the agreement of these observables will be a nontrivial
check of our determination of these backreacted geometries.

According to \cite{Ryu:2006bv,Ryu:2006ef}, 
in a holographic setup the prescription 
to compute entanglement entropy of some region of 
boundary CFT is the following: one takes a closed curve inside of which the region 
of interest lies and then computes the extremal surface 
(with respect to the bulk metric) whose boundary is the given curve. 
Then the entanglement entropy is given by the area of that surface. 
One must proceed with some caution and renormalise the observable,
as the conformal boundary lies 'at infinity' of the bulk space-time and therefore 
any surface reaching conformal boundary has an infinite area.
In the case of various kinds of defects similar calculations of entropy in CFT have already been done
\cite{Bak:2011ga, Azeyanagi:2007qj,Estes:2014hka}. In particular in reference
\cite{Estes:2014hka} entanglement entropy has been obtained for a number 
of known defect configurations.

Since our problem admits translational symmetry along the $y$ axis 
we will only be interested in entanglement entropy contained in a
strip of boundary theory which has the defect in its center.
The first issue one encounters in this calculation is how are curves 
that reach conformal boundary at some given values of the $x$ variable mapped 
under the coordinate transformation connecting Poincar\'e coordinates (\ref{eq:P0}) 
and our slicing ones (\ref{eq:aAds}) for the $T=0$ case, or the transformation given 
by (\ref{eq:BHangleVariable}) for the black hole case. 
The two cases need a separate treatment as the employed coordinate systems
are different.


\subsection{The $T=0$ case}

In the $T=0$  case 
we use the generalized angular coordinate system 
$(r, \alpha)$ with a generic rule of transformation to 
ordinary Fefferman-Graham (FG) patch coordinates 
$z = r f(\alpha)$, $ x = rg(\alpha)$. The functions
$f(\al)$ and $g(\al)$ are no longer simple trigonometric
functions in the backreacted case which poses
some complications in the following.

The problem of finding the relevant extremal surface
reduces to determining the function 
$r(\alpha)$ which solves the equations of motion following
from extremalization of the Nambu-Goto action. 
The entanglement entropy is then determined by evaluation 
of the action on the solution.
With our assumptions NG action takes the following form
\begin{equation}
\mathcal{L}_{\rm NG} = \frac{1}{p\, r(\alpha)^2A(\alpha)^2}\sqrt{r(\alpha)^2+p^2r'(\alpha)^2}~,
\end{equation}
with the resulting equation of motion
\begin{equation}
A(\alpha ) r(\alpha )^2 \left(p^2 r''(\alpha )+r(\alpha )\right)-2 p^2 A'(\alpha ) r'(\alpha ) \left(p^2 r'(\alpha )^2+r(\alpha )^2\right)=0~.
\end{equation}
Since the integral in the Nambu-Goto (NG) action is clearly UV divergent, 
one has to perform a cut-off procedure. In order to combine the 
results both with and without the defect, the regularization has 
to be done covariantly. This is achieved by always expressing 
the  asymptotic part of the metric in FG
coordinates and setting the regulator $z=\epsilon$. This then
translates back to the cutoff $\alpha_c$ in new coordinates
by the solution of $\epsilon=r(\alpha_c)f(\alpha_c)$.
We thus have to determine, to some degree, the coordinate
transformation function $f(\al)$.   

This can be done by writing, in FG coordinates,
the most general form of a metric
which is consistent with the symmetries of the problem
\begin{equation}
ds^2 = \frac{dz^2}{z^2} + C(z/x)\frac{dx^2}{z^2} + D(z/x)\frac{dy^2-dt^2}{z^2}~,
\label{eq:FGmetric}
\end{equation}
Employing now the transformation laws $z = r f(\alpha)$, $ x = rg(\alpha)$
and comparing with our metric ansatz (\ref{eq:Aads}),
one obtains 
\begin{equation}
\frac{f'(\alpha)}{f(\alpha)} = -\frac{\sqrt{1-A(\alpha)^2}}{p A(\alpha)}~,
\hspace{25pt}
\frac{g'(\alpha)}{g(\alpha)} = \frac{A(\alpha)}{p\sqrt{1-A(\alpha)^2}}~.
\label{eq:Transformation}
\end{equation}
The two integration constants are determined by the requirement that 
for $\alpha\rightarrow\pi/2$ we get the empty $AdS$ metric
which boils down to two conditions
\begin{equation}
C(\pi/2) = -\lim_{\alpha\rightarrow\pi/2}\frac{f'(\alpha)f(\alpha)}{g'(\alpha)g(\alpha)}=1~,
\end{equation}
\begin{equation}
D(\pi/2) = \lim_{\alpha\rightarrow\pi/2}\Bigg[C(\alpha)g(\alpha)^2 + f(\alpha)^2\Bigg]=1~.
\end{equation} 
It is important to note, that coordinates (\ref{eq:FGmetric}) do 
not cover the whole spacetime, since $A(0)>1$. 
However, for our purposes, it is enough to consider only the
near boundary region, where formulas work fine.
There, the transformation can be found perturbatively, in parallel to the expansion
of $A(\alpha)$, and it reduces to the ordinary
polar coordinates in the absence of the defect, in which case
it is regular in the whole range of $\alpha$.

To get the entanglement entropy we now must evaluate the NG action
on the solution. Due to translational symmetry along the defect
we quote all formulas per unit length in the $y$ direction.
As it was already mentioned there is a UV divergence
coming from a simple pole of $A(\alpha)$ at $\alpha=\pi/2$. This part
of the integral can be separated and estimated analitically to be
\begin{equation}
S_{\rm div} = \frac{p }{ (\pi/2-\alpha_c)L }~,
\end{equation} 
where $\alpha_c$ is a cut-off defined above by 
$\epsilon=r(\alpha_c)f(\alpha_c)$. To find 
the limiting behaviour we can now expand this equation
near the boundary to obtain 
$\epsilon=Lf'(\pi/2)(\pi/2-\alpha_c)$ to get
the divergent part expressed by a Poincare path
regulator
\begin{equation}
S_{\rm div}= \frac{p f'(\pi/2) }{\epsilon}.
\end{equation}
It turns out that for all our solutions 
$pf'(\pi/2)=1$, which gives the right UV
divergence. 
The regular part follows conformal
invariance so that the total result reads
\begin{equation}
S_{\rm NG} = \frac{1}{\epsilon} - \frac{B}{L}~,
\end{equation}
for some positive $B$ which depends on the amplitude of the source.

%
%

\begin{figure}[t]
\begin{center}
\includegraphics[height = .24\textheight]{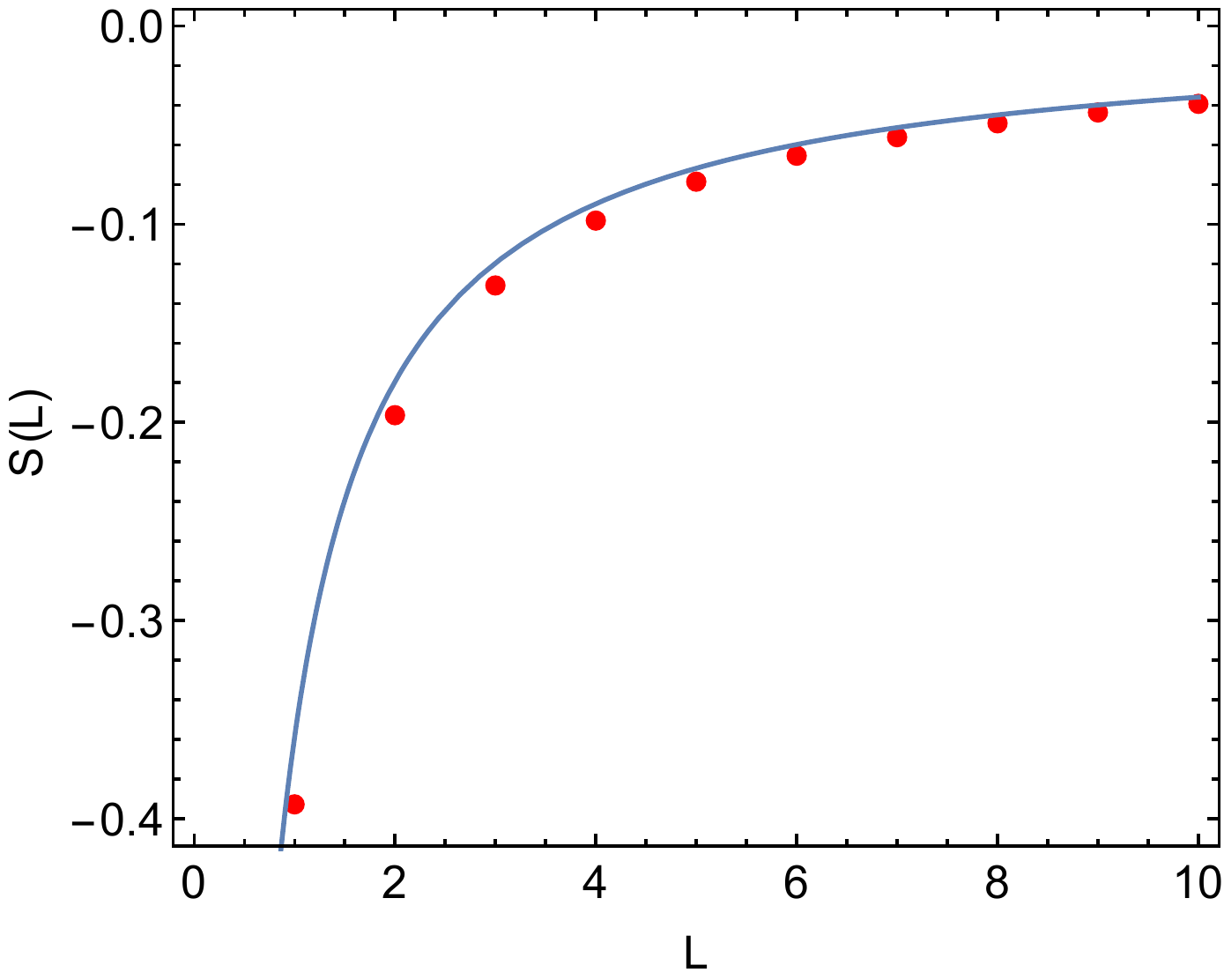} 
\caption{Regular part of the entanglement entropy for the 
defect configuration (dots) and for case of empty $AdS_4$ (line).
Here the defect amplitude corresponds to~$\phi(0)=2$.}
\label{Fig:EET0}
\end{center}
\end{figure}

A surprising feature of the obtained result is that the \emph{regularized} entanglement
entropy is lower than the corresponding one in empty $AdS$ spacetime (see figure \ref{Fig:EET0}).
Indeed for $\phi(0)=1$, a fit to the difference between numerical data and the empty $AdS$ result yields
\eq
S_{\rm defect}-S_{\rm AdS}=\f{-0.0107}{L}~.
\label{e.zeroTEEfit}
\eqx

This is clearly intriguing, although it does not indicate any
contradiction, especially as we are not considering here a different
\emph{state} in the original theory but rather a deformation of the theory
by a nontrivial source added to the Lagrangian.

%
%


\subsection{The finite temperature case}

The entanglement entropy calculation at nonzero temperature follows the same basic steps as before
but with appropriate modifications due to the different coordinate system employed for the 
background geometry.
In this case the corresponding minimal surface will be described
by $(\alpha,y)$ coordinates with a non-trivial dependence $z(\alpha)$. The boundary condition
for the surface is $z'(\pi/2)=-L$ where $L$ is the half width of the strip. In order for the 
equations to be regular we adopt the following ansatz
\begin{equation}
z(\alpha)= Z(\alpha)\cos(\alpha)~,
\label{eq:Ansatz}
\end{equation}  
after which the NG Lagrangian takes the following form
\begin{equation}
\mathcal{L}_{\rm NG} = \frac{\sqrt{S_2(\alpha ,\cos (\alpha ) Z(\alpha ))}}{\cos(\alpha)^2Z(\alpha )^2 \sqrt{1-\cos ^3(\alpha ) Z(\alpha )^3}}
\sqrt{\mathcal{A}}~,
\end{equation}
where
\begin{eqnarray}
\mathcal{A} &=& \left(1-\cos ^3(\alpha ) Z(\alpha )^3\right) S_1(\alpha ,\cos (\alpha ) Z(\alpha )) \Bigg(Z'(\alpha ) (\cos (\alpha ) F(\alpha ,\cos
   (\alpha ) Z(\alpha ))\\
	\nonumber
	&+&\sin (\alpha ))+Z(\alpha ) (\cos (\alpha )-\sin (\alpha ) F(\alpha ,\cos (\alpha ) Z(\alpha
   )))\Bigg)^2+\\
	\nonumber
	&+&H_2(\alpha ,\cos (\alpha ) Z(\alpha )) \Bigg(\sin (\alpha ) Z(\alpha )-
	\cos (\alpha ) Z'(\alpha )\Bigg)^2~.
\label{eq:A}
\end{eqnarray}
The equation of motion
is a non-linear ordinary differential equation defined
numerically in terms of the profiles calculated in section~\ref{solTnon0}. 
The boundary conditions for the extremal surface translate to: $Z'(0)=0$ and $Z(\pi/2)=L$.
The resulting equation with these boundary conditions can be solved using
spectral discretization with the Newton method.

\begin{figure}[h!]
\begin{center}
\includegraphics[height = .214\textheight]{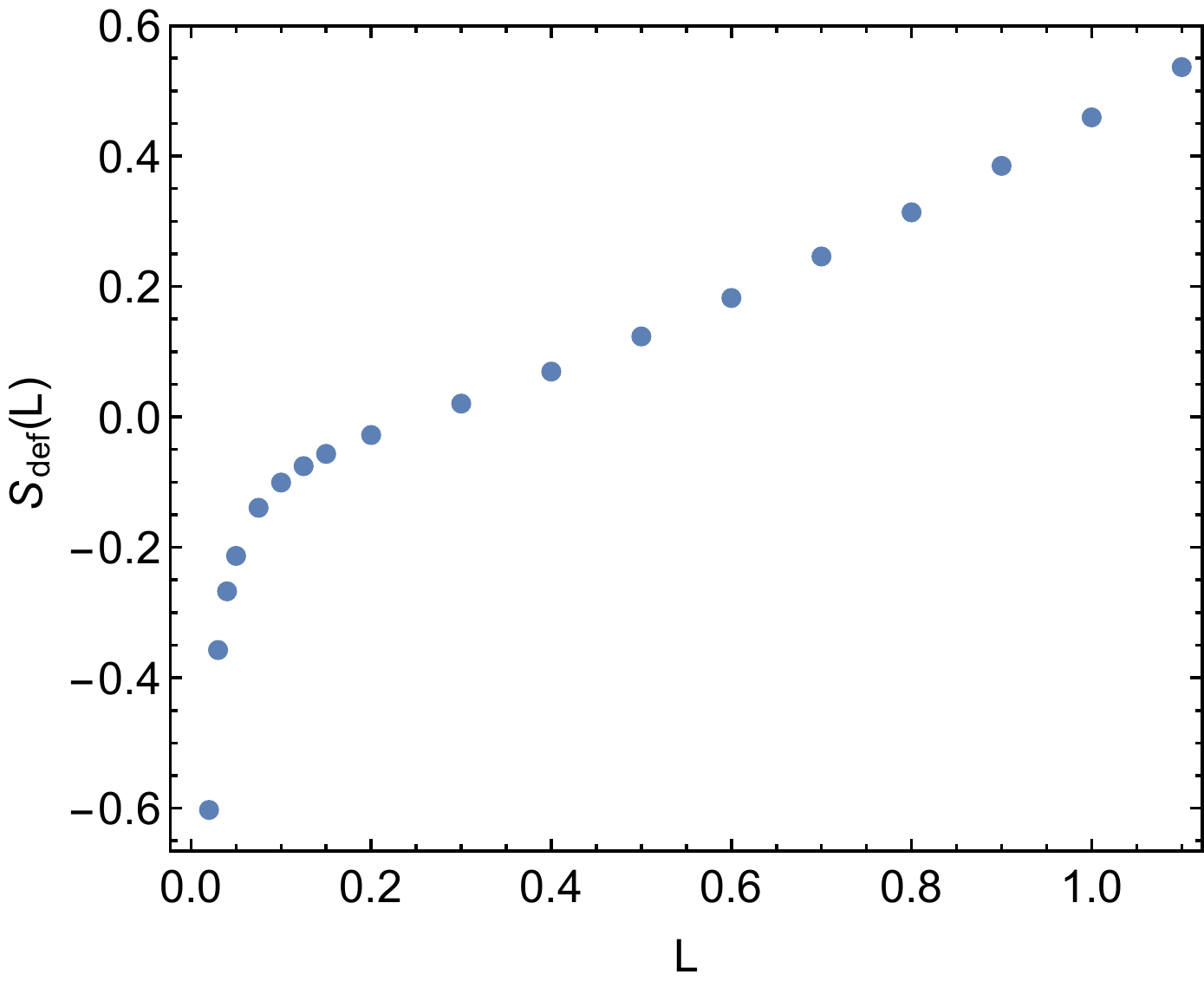} 
\includegraphics[height = .21\textheight]{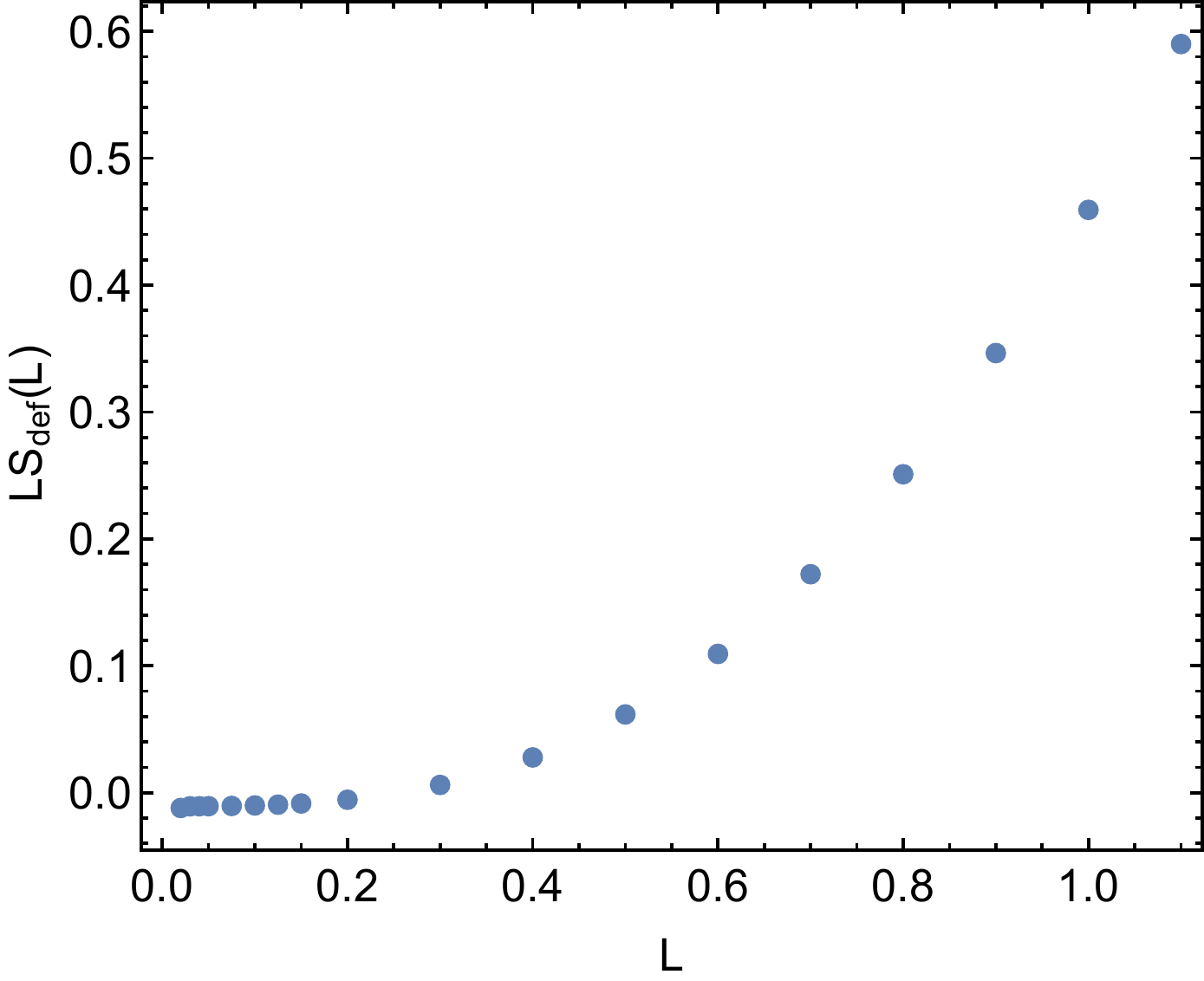}
\caption{Entanglement entropy for a Dirac delta defect with $\phi(0)=1.0$ at finite
temperature
with $N=50$ spectral grid.}
\label{Fig:EETnon0}
\end{center}
\end{figure}

In order to evaluate the entanglement entropy we regularize
the divergent action, evaluated on-shell, by subtracting pointwise
the proper $AdS$ NG Lagrangian evaluated on a corresponding extremal surface in empty $AdS$,
found numerically in an analogous coordinate system on the same numerical grid in $\al$ i.e.
\begin{equation}
S_{\rm def}(L) = \int_0^{\pi/2} d\alpha(\mathcal{L}_{\rm NG} - \mathcal{L}_{\rm AdS})~.
\label{eq:DL}
\end{equation}
We calculate the entanglement entropy for a Dirac delta defect
characterized by $\phi(0)=1$ for the (half-)width of the strip the range $0.02\leq L\leq1.1$.
The results are shown in figure~\ref{Fig:EETnon0}.


\begin{figure}[h!]
\begin{center}
\includegraphics[height = .214\textheight]{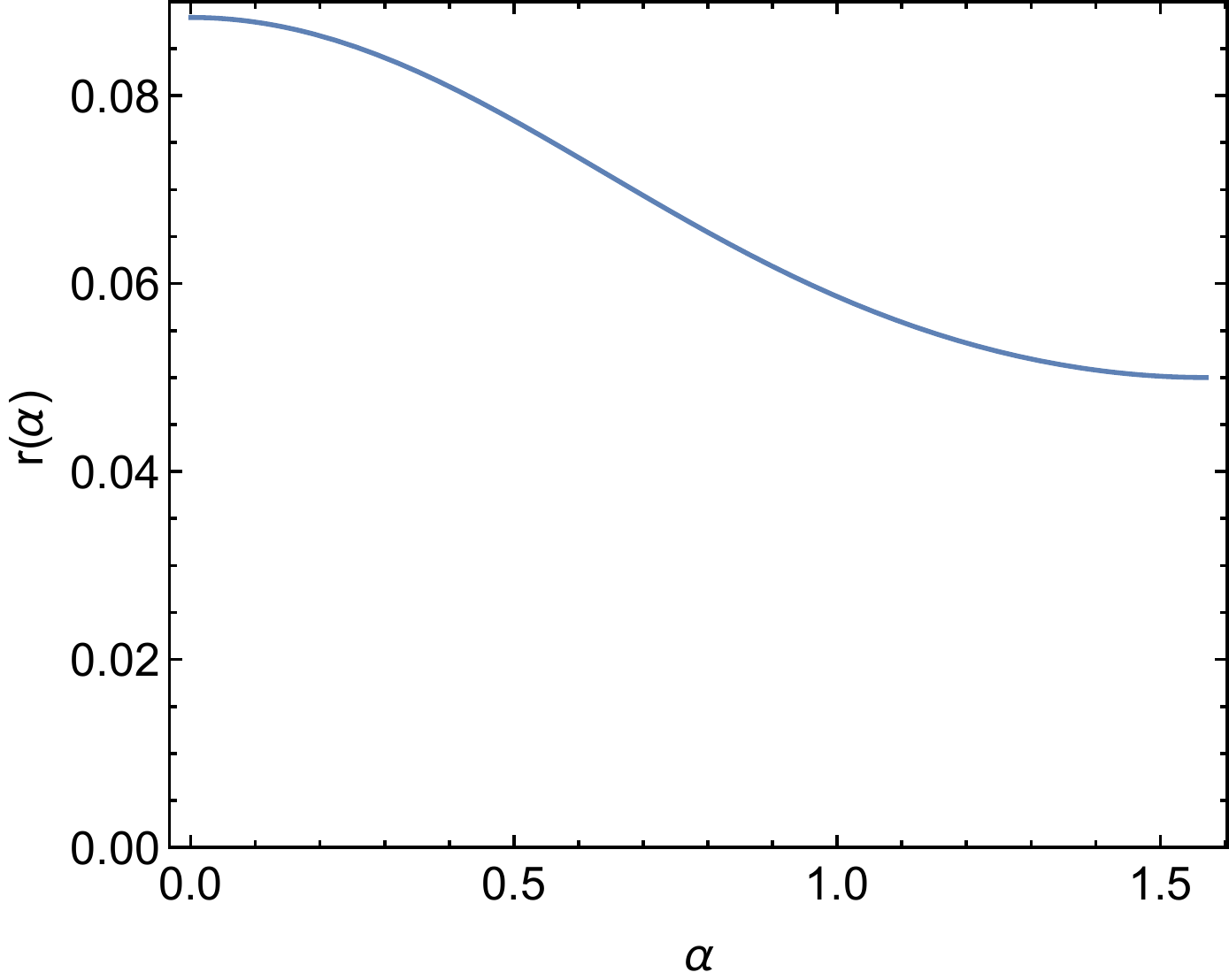} 
\includegraphics[height = .21\textheight]{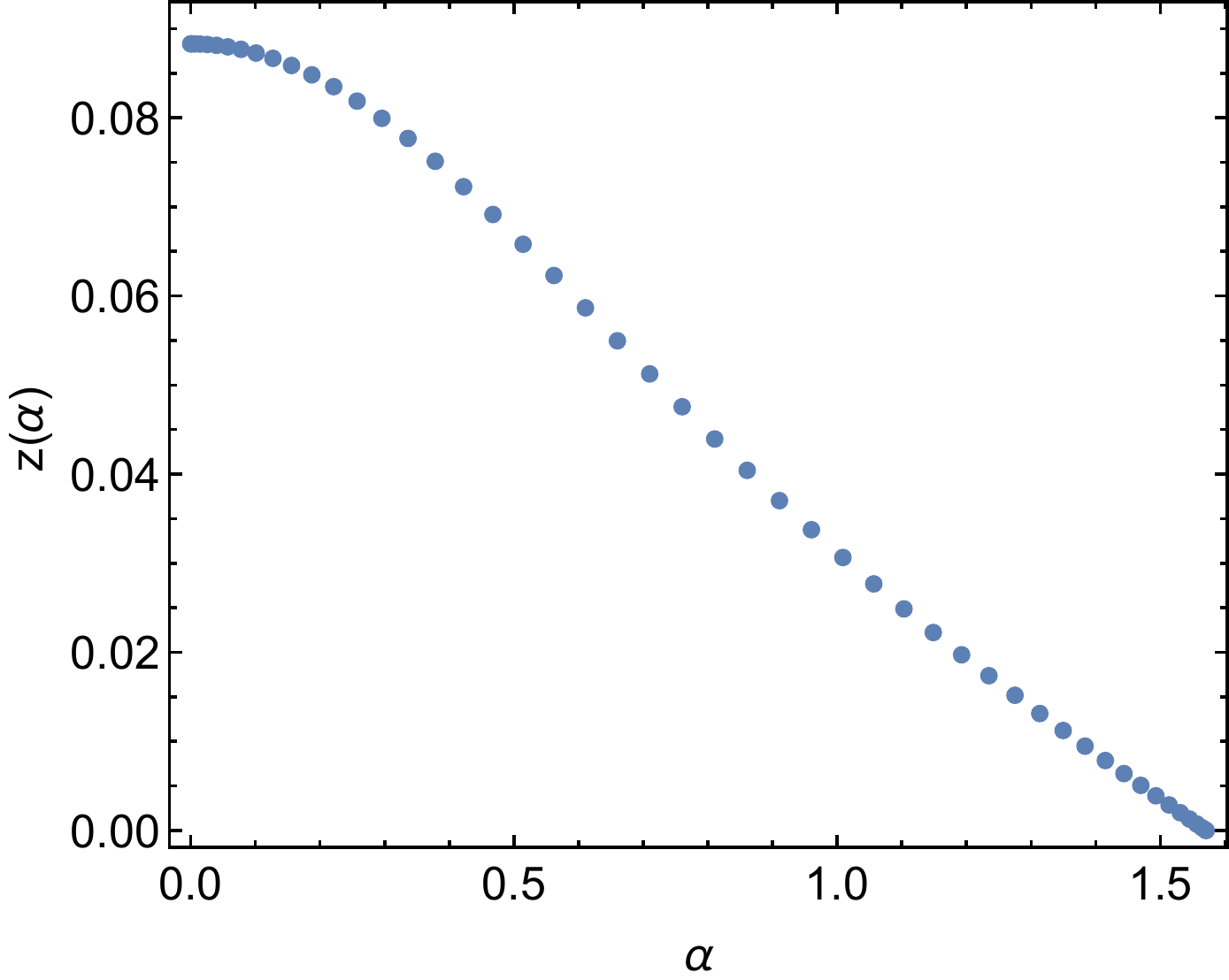}
\caption{Plots of the minimal surfaces embeddings for $L=0.05$
for $T=0$ (left panel) and $T>0$ (right panel).}
\label{Fig:EETnon0emb}
\end{center}
\end{figure}

Due to the conformal symmetry of the defect the generic form of 
the entanglement entropy has to be
\begin{equation}
S_{\rm def}(L) =  \frac{B(LT)}{L}~,
\label{eq:GenericEE}
\end{equation}
where $B(LT)$ is some smooth function. Of course $B(LT)$ will also
depend on the (dimensionless) amplitude of the Dirac delta function.
This formula allows for making a connection with the entanglement entropy
computation for $T=0$ presented earlier.
Indeed, for fixed $T$ when 
$L\rightarrow0$ we recover the vacuum defect result.
For this particular case the $1/L$ falloff has a coefficient
$a=-0.0107$ which is consistently checked against zero
temperature calculation (\ref{e.zeroTEEfit}).
This is also a nontrivial check of our numerical constructions of the relevant
backgrounds, as due to the different coordinate systems employed the extremal
surfaces for small $L$ obtained in both cases are quite different as can be seen
on Fig.~\ref{Fig:EETnon0emb}.


\begin{figure}[t]
\begin{center}
\includegraphics[height = .24\textheight]{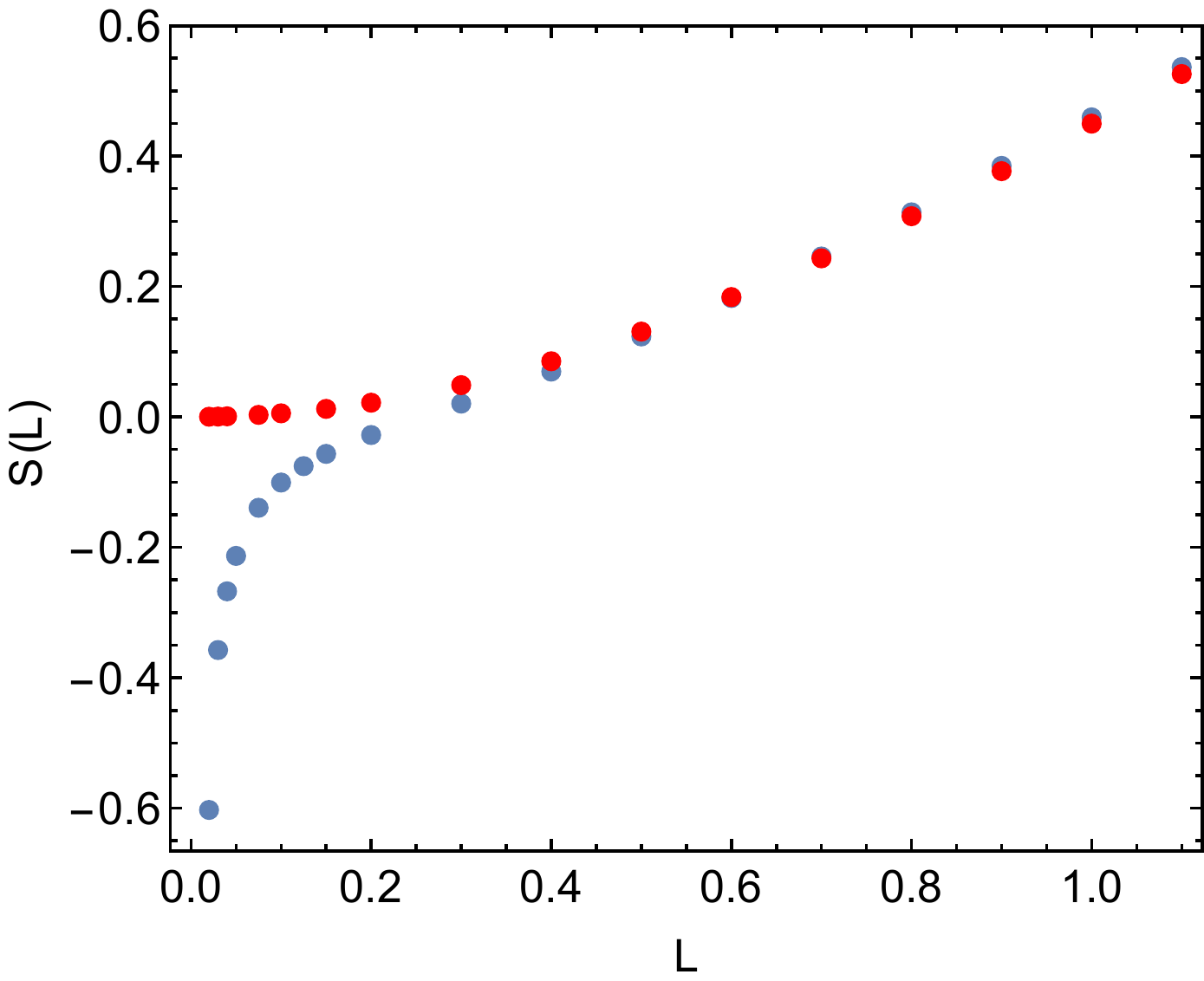} 
\caption{Regular part of the entanglement entropy for the 
defect configuration (blue dots) and for $AdS$ black hole case (red dots).}
\label{EETD}
\end{center}
\end{figure}

It is quite interesting to compare the finite temperature entanglement entropy
with and without the defect. The relevant plots are shown in figure~\ref{EETD}.
Again we observe the rather surprising result that the defect source
decreases the entanglement entropy. Clearly, for small $L$
the effects of the defect are significant while for $L>0.25$ thermal effects
start to dominate.


\section{Conclusions}
\label{concl}

In this paper we constructed holographic duals of strongly coupled three-dimensional CFT's deformed by a localized
Dirac delta source. The motivation for this work was to move towards a holographic construction
mimicking a crystalline lattice with pointlike localized sources. In the present paper we have
concentrated on developing the necessary numerical relativity methods in order to consistently
handle Dirac delta like sources and considered explicitly a single defect along a line both at
zero and at nonzero temperature.

The 1D Dirac delta source for a scalar operator of dimension $\Dl=2$ is scale invariant.
We found that requiring that the backreacted geometry respects this scale invariance
imposes very stringent constraints on the scalar potential which is consequently uniquely determined.
An intriguing outcome is that the resulting potential is exactly the scalar potential
appearing in certain Kaluza-Klein reductions of 11D supergravity. All further considerations
in the present paper were performed with this concrete choice of the scalar self-interaction.

In order to find the dual backgrounds, we had to use systems of coordinates which were
adapted to the presence of the Dirac delta source and which took into account 
the high variability of the scalar field and of the metric coefficients in the infinitesimal
neighbourhood of the point of insertion of the Dirac delta function on the boundary.

At zero temperature, we constructed the dual backgrounds in two ways: using a perturbative
expansion and performing a direct numerical solution of the equations of motion using
an $AdS$ slicing analogous to the one used for obtaining the Janus solution
\cite{Bak:2003jk,DHoker:2009gg}.

For nonzero temperature, we adopted the DeTurck method which required, however,
two modifications. Firstly, the adapted choice of coordinates (similar but different
from the one that we used for $T=0$) was encoded in the choice of coordinates used
for the reference $AdS$ black hole metric. Secondly, the boundary at $z=0$ of the numerical
grid represented really the infinitesimal neighbourhood of the Dirac delta source
and the values of the fields there had to be determined from the equations of motion.
With those two modifications in place we constructed the numerical background and performed two
cross checks. We compared the resulting numerical background with analogous geometries obtained for 
regularized delta sources. We also compared the small size limit of entanglement entropy with the entanglement
entropy evaluated at zero temperature.

Incidentally, we found that the entanglement entropy evaluated in the theory deformed by
the Dirac delta source is \emph{lower} than the analogous quantity without the defect.
It would be very interesting to understand this property from a more physical perspective.

There are numerous directions for further study like the construction of a lattice
of Dirac delta defects in order to study phenomena analogous to \cite{Horowitz:2012ky},
extensions to chemical potentials and pointlike sources. It would be interesting to determine
whether the `hovering black hole' phenomena observed in \cite{Horowitz:2014gva}
have their counterpart in the present context. We intend to address at least some 
of these issues in subsequent work.

\vspace{15pt}
\noindent {\bf Acknowledgements:} The authors would like to thank
Maciej Maliborski, Hesam Soltanpanahi, Pawe{\l} Caputa, Ludwik Turko, Jorge Santos and Przemys\l aw Witaszczyk for
discussions.  RJ and JJ wish to thank Galileo
Galilei Institute for Theoretical Physics for hospitality and the INFN for
partial support during the program \emph{Holographic Methods for Strongly
  Coupled Systems} where this work was finalized. RJ was partially supported by
the Polish National Science
Center (NCN) grant 2012/06/A/ST2/00396 while JJ by the NCN post-doctoral internship grant DEC-2013/08/S/ST2/00547.

\bibliography{defectbiblio}{}
\bibliographystyle{utphys}

\end{document}